\documentclass[a4paper,12pt]{article}

\usepackage[left=2.5cm,right=2.5cm,top=2.5cm,bottom=2.5cm]{geometry}
\usepackage{graphicx}
\usepackage{amssymb}
\usepackage{multirow}
\usepackage{amsmath}
\usepackage{psfrag}
\usepackage{setspace}
\usepackage{subcaption}
\usepackage[colorlinks=true,bookmarks=true]{hyperref}
\usepackage{float}
\usepackage{tabularx}
\usepackage{array}
\usepackage{cite}
\usepackage[utf8]{inputenc}
\usepackage{authblk}

\hypersetup{citecolor=blue}
\newcommand{\dif}{\mathrm{d}}
\newcommand{\Eqref}[1]{(\ref{#1})}
\newcommand{\half}{\frac{1}{2}}

\newcommand{\brac}[1]{\left(#1 \right)}
\newcommand{\sbrac}[1]{\left[#1\right]}

\begin{document}

\title{Periodic orbits around a spherically symmetric naked singularity}
\author[1]{Gulmina Zaman Babar\thanks{gulminazamanbabar@yahoo.com}}
\author[2]{Adil Zaman Babar\thanks{adilzamanbabar@gmail.com}}
\author[3]{Yen-Kheng Lim\thanks{phylyk@nus.edu.sg}}

\affil[1]{\normalsize{\textit{School of Natural Sciences, National University of Sciences and Technology, Sector H-12, Islamabad, Pakistan}}}
\affil[2]{\normalsize{\textit{Department of Electrical Engineering, University of Engineering and Technology, Peshawar, Pakistan}}}
\affil[3]{\normalsize{\textit{Department of Physics, National University of Singapore, 117551 Singapore, Singapore}}}

\date{\normalsize{\today}}
\maketitle

\begin{abstract}
  The motion of time-like test particles in the Fisher/Janis-Newman-Winicour (F/JNW) spacetime is studied with the Hamiltonian formulation of the geodesic equations. The spacetime is characterised by its mass parameter $r_g$ and scalar field parameter $\nu$. The innermost bound and stable circular orbits are calculated and the effective potential is analysed. Consistent with numerical results in earlier literature, for $\nu<\half$, particles with non-zero angular momentum encounter an infinite potential barrier, preventing them from reaching the naked singularity at $r=r_g$. Periodic orbits in the spacetime are also obtained. Compared to the periodic orbits around the Schwarzschild black hole, it is found that typically lower energies are required for the same orbits in the F/JNW spacetime.
\end{abstract}

\section{Introduction} \label{intro}

One of the most notable features where General Relativity departs from Newtonian gravity is the precession of orbital trajectories. The explanation of Mercury's perihelion precession which was previously unaccounted for by Newtonian calculations was famously one of the three observational tests which firmly established the grounds for General Relativity.

More recently, studies of relativistic orbits are of interest again especially with the recent detection of gravitational waves. These waves are generated by inspiralling black-hole binaries, and thus their signature might depend on their relativistic trajectories. In light of this, Levin et al., proposed a classification system \cite{Levin:2008mq,Levin:2008ci,Levin:2009sk} of relativistic trajectories around a gravitating source to fully understand its dynamics, which are particularly relevant in the case inspirals of bodies with extreme mass ratios. Their classification scheme follows the paradigm by Poincar\'{e}  which states that the behaviour of a dynamical system can be understood through a study of its periodic trajectories. Hence Levin and Perez-Giz sought to classify periodic orbits around a Kerr black hole by three integers: $z$, $w$, and $v$ for the zoom, whirl, and vertices, respectively.

The Laser Interferometer Gravitational-wave Observatory (LIGO) detectors recently detected the gravitational waves generated as a result of two coalescing black holes \cite{Abbott:2016blz}. Reference \cite{Apostolatos:1994mx} provides a constructive analysis of the inspiralling compact binaries causing the gravitational waves as a result of spin-induced orbital precession. Gravitational radiation produced by the eccentric orbits with a zoom-whirl behaviour were studied in Ref.~\cite{Glampedakis:2002ya}. A complete classification of the zoom-whirl structure of the periodic orbits was done in Refs.~\cite{Levin:2008mq,Levin:2008ci,Levin:2009sk}. The analysis of null and massive test-particle motion was carried out in \cite{Hod:2012ax,Hackmann:2008tu,Hackmann:2010zz,Fujita:2009bp}.

On a different front, research from the last couple of decades has unveiled the fact that the end state of complete gravitational collapse could result in the evolution of naked singularities \cite{Eardley:1978tr,Christodoulou:1984mz,Joshi:2001xi,Joshi:2013dva}. Accretion properties and gravitational lensing were used in Refs.~\cite{Virbhadra:2002ju,Gyulchev:2008ff} to distinguish black holes from a naked singularity. The prospect of the possible existence of a naked singularity was recently proposed by Crisford and Santos \cite{Crisford:2017zpi}, where it was shown that the violation of the weak cosmic censorship conjecture was possible in four-dimensional Einstein-Maxwell gravity with a cosmological constant. In light of these possibilities, our aim will be to explore the physical phenomena in the vicinity of these naked singularities to compare with those occurring in the well-studied case of black holes with regular event horizons.

In this paper, we will be interested in the motion of particles around a spherically symmetric naked singularity. The simplest model for such a problem would be an extension of Schwarzschild geometry by dressing it with a massless scalar field. This particular spacetime has had a long history and was known under various names. It is perhaps most widely known as the Janis-Newman-Winicour \cite{Janis:1968zz} solution, though it turns out to be a rediscovery of a solution obtained by Fisher \cite{Fisher:1948yn}. For simplicity we shall refer to it as the Fisher/Janis-Newman-Winicour (F/JNW) solution. The physical properties of the solution were further explored by Janis, Robinson, and Winicour \cite{Janis:1968zz} and hence some papers have referred to it as the Fisher-Janis-Robinson-Winicour solution. More recently, it was discovered \cite{Virbhadra:1997ie} that what was previously thought to be a separate solution by Wyman \cite{Wyman:1981bd} was actually identical to the F/JNW solution. We will use specific features of the periodic orbits to distinguish the Schwarzschild black hole from the naked singularity. Various aspects of particle motion in the F/JNW spacetime have been previously considered in \cite{Zhou:2014jja,Kovacs:2010xm,Chowdhury:2011aa,Babar:2015kaa}. In particular, in this paper we consider periodic orbits with rational orbital frequencies.

This paper is organised as follows. In Sec.~\ref{metric}, we obtain the geodesic equations in the F/JNW spacetime using the Hamiltonian formulation. Sec.~\ref{EffPot} encompasses the derivation of the innermost bound and stable circular orbits, as well as a qualitative analysis of the effective potential. In Sec.~\ref{periodic} the existence of periodic orbits in the F/JNW is studied. This paper closes with some concluding remarks in Sec.~\ref{conclusion}.

\section{Geodesic equations} \label{metric}

We are considering a test particle moving in the Fisher/Janis-Newman-Winicour spacetime \cite{Fisher:1948yn,Janis:1968zz}, given by
\begin{align}
 \dif s^2&=-f^\nu\dif t^2+f^{-\nu}\dif r^2+r^2f^{1-\nu}\brac{\dif\theta^2+\sin^2\theta\,\dif\phi^2},\\
   f&=1-\frac{r_g}{r}. \label{JNWmetric}
\end{align}
This metric, along with the massless scalar
\begin{align}
 \varphi&=\sqrt{\frac{1-\nu^2}{2}}\ln f, \label{JNWscalar}
\end{align}
is a solution to Einstein gravity minimally coupled to a massless scalar.
\par
The motion of the test particle is described by a trajectory $x^\mu(\tau)$, where $\tau$ is an appropriate affine parametrisation. In this paper we shall use the Hamiltonian description to derive the equations of motion. To this end we require the conjugate momenta $p_\mu=g_{\mu\nu}\dot{x}^\nu$, where the components are explicitly given by
\begin{align}
 p_t&=-f^\nu\dot{t},\quad p_r=f^{-\nu}\dot{r},\quad p_\theta=r^2f^{1-\nu}\dot{\theta},\quad p_\phi=r^2f^{1-\nu}\sin^2\theta\,\dot{\phi}. \label{momenta}
\end{align}
The Hamiltonian, $H=\half g^{\mu\nu}p_\mu p_\nu$ for this background is
\begin{align}
 H&=\half\brac{-f^{-\nu}p_t^2+f^\nu p_r^2+\frac{p_\theta^2}{r^2f^{1-\nu}}+\frac{p_\phi^2}{r^2f^{1-\nu}\sin^2\theta}}.
\end{align}
This Hamiltonian is a constant of motion, where its value can be fixed by a normalisation condition $H=\half g^{\mu\nu}p_\mu p_\nu=\half\epsilon$, where $\epsilon=-1$ for time-like geodesics and $\epsilon=0$ for null geodesics.

Since the Hamiltonian is cyclic in $t$ and $\phi$, the momenta along these directions are conserved and therefore we have
\begin{align}
 \dot{p}_t&=0,\quad\dot{t}=f^{-\nu}E,\\
 \dot{p}_\phi&=0,\quad\dot{\phi}=\frac{L}{r^2f^{1-\nu}\sin^2\theta},
\end{align}
where $E$ and $L$ are constants of motion which we regard, respectively, as the energy and angular momentum of the particle. Applying the Hamilton equations for the remaining two coordinates gives
\begin{align}
 \dot{r}&=f^\nu p_r,\\
 \dot{p}_r&=\half\sbrac{-p_r^2\partial_r\brac{f^\nu}-p_\theta^2\partial_r\brac{r^{-2}f^{-(1-\nu)}}+\partial_r\brac{f^{-\nu} R}+\Theta\partial_r\brac{r^{-2}f^{-(1-\nu)}}},\\
 \dot{\theta}&=r^{-2}f^{-(1-\nu)}p_\theta,\\
 \dot{p}_\theta&=\half r^{-2}f^{-(1-\nu)}\partial_\theta\Theta,
\end{align}
where
\begin{align}
 R&=E^2-f^\nu\brac{1+\frac{L^2}{r^2f^{1-\nu}}}-\frac{Q}{r^2f^{1-2\nu}},\\
 \Theta&=Q-\frac{L^2\cos^2\theta}{\sin^2\theta},
\end{align}
and $Q$ is another separation constant, which will be zero in our case of a spherically symmetric spacetime.
\par
Since this spacetime is spherically symmetric, we can always find a coordinate transformation that places the trajectory on the equatorial plane. As such, we may consider without loss of generality $\theta=\frac{\pi}{2}$ and $Q=0$. Since we are considering time-like particles in this paper, we will also set $\epsilon=-1$.

\section{Effective potential and circular orbits}\label{EffPot}

The Hamiltonian constraint $g^{\mu\nu}p_\mu p_\nu=\epsilon$ leads to the first-order equation which allows the motion to be described in terms of an effective potential $U^2$, where
\begin{align}
 \dot{r}^2&=E^2-U^2,\quad U^2=\frac{L^2}{r^2}f^{2\nu-1}+f^\nu.  \label{U_eff}
\end{align}
We can clearly see that $U^2\rightarrow 1$ as $r\rightarrow\infty$, as expected for an asymptotically flat spacetime. Since $U^2\rightarrow 1$ as $r\rightarrow \infty$, particles with $E>1$ are able to escape to infinity. Thus $E=1$ is the demarcation between bound and unbound orbits.
\par
\begin{figure} 
 \begin{center}
  \includegraphics{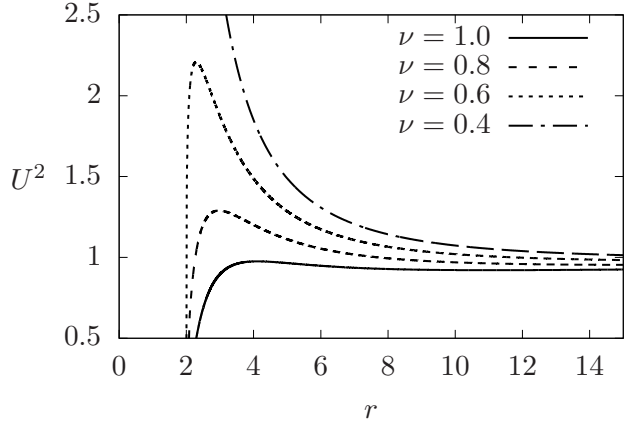}
  \caption{Effective potential for $r_g=2$, $L=3.9$, and various $\nu$.}
  \label{fig_EffPotential1}
 \end{center}
\end{figure}

The term $\frac{L^2}{r^2}f^{2\nu-1}$ in the effective potential gives us interesting behaviours close to the singularity at $r=r_g$. If $\nu<\half$ and $L\neq 0$, the potential diverges as $r\rightarrow r_g^+$, as was noted by earlier works \cite{Zhou:2014jja,Chowdhury:2011aa}. This infinite barrier vanishes if $L=0$. On the other hand, if $\nu>\half$, the potential approaches zero for any $L$, which is a behaviour similar to the well-known Schwarzschild geodesics. Figure \ref{fig_EffPotential1} shows a typical structure of the effective potentials for decreasing $\nu$. We see that as $\nu<\half$, the potential barrier turns into an infinite one.
\par
Trajectories with constant $\dot{r}=0$ describe circular orbits. Particles will remain at constant $r$ if they sit at an equilibrium position defined by $\frac{\dif\brac{U^2}}{\dif r}=0$. For an equilibrium position $r$, the angular momentum and energy required for a circular orbit is
\begin{align}
 L&=r\sqrt{\frac{\nu r_g\brac{1-r_g/r}^{1-\nu}}{2r-(1+2\nu)r_g}},\quad E=\sqrt{\brac{1-\frac{r_g}{r}}^\nu\brac{\frac{2r-r_g(1+\nu)}{2r-r_g(1+2\nu)}}}. \label{Circular_EL}
\end{align}
The denominators of Eq.~\Eqref{Circular_EL} are real if and only if $2r-r_g(1+2\nu)\geq0$. The limiting case of an equality is
\begin{align}
 r_{\mathrm{ph}}=\half r_g(1+2\nu),
\end{align}
which corresponds to a circular photon orbit of $\epsilon=0$.

\begin{figure} 
 \begin{center}
  \includegraphics{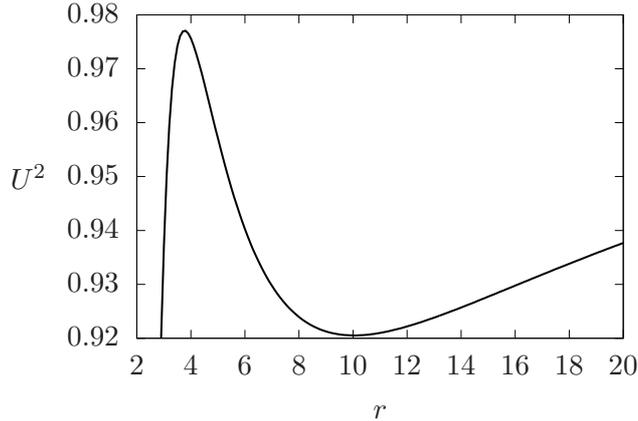}
  \caption{Effective potential for a time-like particle with $L=3.5$ around a spacetime with parameters $r_g=2$ and $\nu=0.9$.}
  \label{fig_PotExtrema}
 \end{center}
\end{figure}

For a given angular momentum $L$, the radius of the circular orbit can be determined using Eq.~\Eqref{Circular_EL}. The stabilities of these orbits are determined by $\frac{\dif^2\brac{U^2}}{\dif r^2}$ evaluated at the orbital radius. Stable orbits correspond to $\frac{\dif^2\brac{U^2}}{\dif r^2}<0$, and unstable ones have $\frac{\dif^2\brac{U^2}}{\dif r^2}>0$. It is easy to see that the unstable and stable orbits are two successive extrema of the effective potential $U^2$. For example Fig.~\ref{fig_PotExtrema} shows the effective potential of a particle with angular momentum $L=3.5$ around a naked singularity of $r_g=2$ and $\nu=0.9$. In this case, the unstable circular orbit is located at $r=3.766$ and the stable one is at $r=10.033$. By tuning the angular momentum, the two circular orbits will merge into a single value where $\frac{\dif^2\brac{U^2}}{\dif r^2}=0$. We will denote this case as \emph{marginally stable circular orbits}, and they are given by
\begin{align}
 r_{\mathrm{ms}}&=\half r_g\brac{1+3\nu\pm\sqrt{5\nu^2-1}}.\label{r_ms}
\end{align}

Note that for $\nu>\half$, the solution with the lower sign gives $r_{\mathrm{ms}}<r_g$ and is not of interest. However, for the range $\frac{1}{\sqrt{5}}\leq\nu\leq\half$, both roots lie outside the singularity with $r_{\mathrm{ms}}>r_g$. Recalling that marginally stable circular orbits result from a merging of two circular orbits, it follows that having two marginally stable orbits for the range $\frac{1}{\sqrt{5}}\leq\nu\leq\half$ implies that there are at least three circular orbits for an appropriate choice of $L$ \cite{Chowdhury:2011aa}, corresponding to two potential wells.

For the rest of the paper, we shall be interested in the range $\nu\geq\half$, so that we have two circular orbits outside the naked singularity, with the minimum angular momentum for bound orbits $L_{\mathrm{ms}}$ which is obtained by substituting the upper root of \Eqref{r_ms} into \Eqref{Circular_EL}.

In this paper, we are interested in periodic orbits which are confined within a finite region where $E^2\geq U^2$. The boundaries of the motion occur at the turning points $\dot{r}=0$, which, according to Eq.~\Eqref{U_eff} can be found by setting $E^2=U^2$ and solving for $r$. Thus, a bound orbit is characterised by the particle satisfying $\dot{r}^2\geq 0$ within the range $r_-\leq r\leq r_+$, where $r_\pm$ are the roots of $E^2=U^2$.

In order to have a sufficiently rich variety of periodic orbits, the potential well must be sufficiently deep so that bound states may exist for particle with a wide range of $E<1$. By inspection of the qualitative features of \Eqref{U_eff}, we see that decreasing $\nu$ at fixed $L$ raises the potential barrier and makes the well more shallow, decreasing the number of possible periodic orbits. Thus, for a given $\nu$, an appropriate range of $L$ must be chosen in order to have a sufficiently deep well that supports a wider variety of orbits.

Since the notion of a `sufficiently deep' well is ill-defined, we shall provisionally define a parameter $L_{\mathrm{mb}}$, the value of angular momentum where the peak of the potential barrier coincides with $U^2=1$. A circular orbit with $E=1$ on the peak of the well is unstable, and we denote this as a \emph{marginally bound} orbit. A small perturbation about this marginally bound orbit will send the particle either towards the $r=r_g$ singularity, or asymptotically towards infinity at $r\rightarrow\infty$.

For each $\nu$, we may uniquely determine $L_{\mathrm{mb}}$ and $L_{\mathrm{ms}}$. If we further consider
\begin{align}
 L_{\mathrm{av}}=\half\brac{L_{\mathrm{ms}}+L_{\mathrm{mb}}}, \label{L_av}
\end{align}
this would give an appropriate potential well for any $\nu$ that captures most of the physics of the bound orbits. In the Schwarzschild case $\nu=1$ and $r_g=2$, we find that $L_{\mathrm{av}}\simeq3.732$, which is close to the angular momentum values chosen in Ref.~\cite{Levin:2008mq} to depict periodic orbits.
\begin{figure} 
 \begin{center}
  \includegraphics{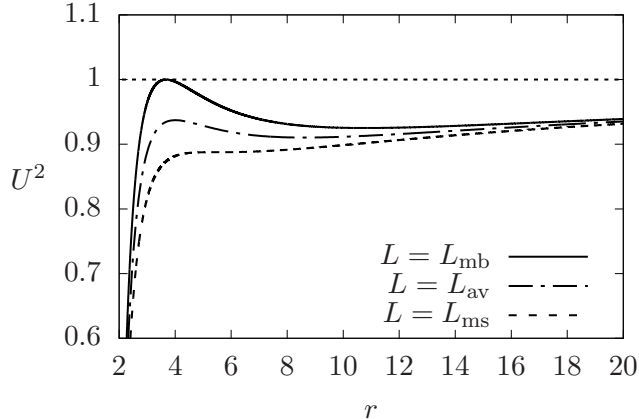}
  \caption{Effective potentials with angular momenta $L_{\mathrm{mb}}$, $L_{\mathrm{ms}}$, and $L_{\mathrm{av}}$ for $r_g=2$ and $\nu=0.9$.}
  \label{fig_EffPotential2}
 \end{center}
\end{figure}

\begin{figure} 
 \begin{center}
  \includegraphics{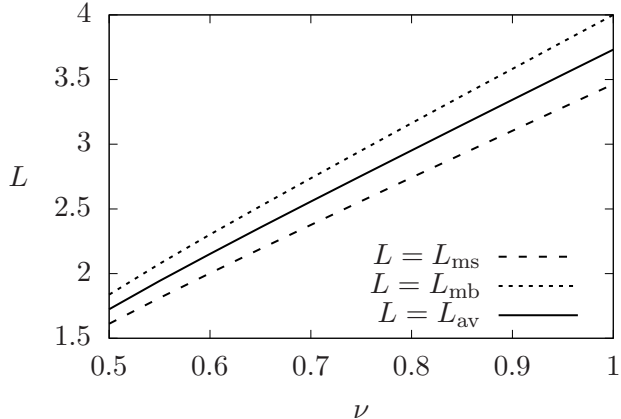}
  \caption{Values of $L_{\mathrm{mb}}$, $L_{\mathrm{ms}}$, and $L_{\mathrm{av}}$ for various $\nu$.}
  \label{fig_Lrange}
 \end{center}
\end{figure}
Figure \ref{fig_EffPotential2} shows an example of the potentials for $L_{\mathrm{mb}}$, $L_{\mathrm{ms}}$, and $L_{\mathrm{av}}$ for the case $r_g=2$ and $\nu=0.9$. The values of these angular momenta for $\half\leq\nu\leq 1$ are depicted in Fig.~\ref{fig_Lrange}, where we see that for lower values of $\nu$, the typical range of angular momentum is decreased to have a sufficiently deep potential well.

\section{Periodic orbits}\label{periodic}

In this section, we seek periodic time-like orbits around the F/JNW spacetime. As argued in Refs.~\cite{Levin:2008mq,Levin:2009sk}, all orbits appear like perturbations of periodic orbits. Hence, an understanding of the structure of periodic orbits allows us to understand the complete nature of any generic orbits within the spacetime. In this section, we shall establish that periodic orbits in the F/JNW spacetime have the same qualitative features as the Schwarzschild spacetime, albeit with different particle energies.

To begin, we shall briefly review the methods of Levin et al. \cite{Levin:2008mq,Levin:2009sk} to study the periodic orbits. Any bound orbit may be characterised by its frequency of oscillations in its radial coordinate $r$ and its angular coordinate $\phi$. A periodic orbit corresponds to trajectories where the ratio of these two frequencies is a rational number, such that the trajectory will close and the particle returns to its initial state within a finite (affine) time, thus executing its prior trajectory repeatedly.

Loosely speaking, we may define the \emph{radial} period of the particle as the (affine) time taken to return to the same radius upon starting at the apastron  $r_+$ (where $E^2=U^2$) to return to the same radius. If within this time, the evolution of $\phi$ is an integer multiple of $2\pi$, we have a periodic orbit. Therefore, the quantity of interest here is
\begin{align}
 \Delta\phi=2\int_{r_-}^{r_+}\frac{\dot{\phi}}{\dot{r}}\dif r=2\int_{r_-}^{r_+}\frac{L\,\dif r}{r^2f^{1-\nu}\sqrt{E^2-\brac{\frac{L^2}{r^2}f^{2\nu-1}+f^\nu}}}.\label{DeltaPhiIntegral}
\end{align}
Adopting the terminology of Ref.~\cite{Levin:2008mq}, we define $q$ by
\begin{align}
 q=\frac{\Delta\phi}{2\pi}-1.
\end{align}
Periodic orbits correspond to orbits where $q$ is a rational number. Generically, the quantity $q$ is interpreted as the amount of precession beyond a closed ellipse \cite{Levin:2009sk}.

For a given $\nu$ and $r_g$, the value of $q$ may be regarded as a function of $E$ and $L$. This is because a given $E$ and $L$ fixes the values of $r_\pm$, and thus fixes the integral in Eq.~\Eqref{DeltaPhiIntegral}. If we further consider the case $L=L_{\mathrm{av}}$, where $L_{\mathrm{av}}$ are fixed by $\nu$ and $r_g$, then $q$ depends only on $E$. For the case $\nu=1$, the integral in Eq.~\Eqref{DeltaPhiIntegral} can be given in a closed-form expression \cite{Chandrasekhar:1985kt,Levin:2008mq,Levin:2009sk} and hence $q$ can be evaluated exactly. For general $\nu$, we calculate the values of $q$ numerically.

If $q$ is a rational number, it may be decomposed in terms of three integers $(z,w,v)$, where
\begin{align}
 q=w+\frac{v}{z}.
\end{align}
As shown in Ref.~\cite{Levin:2008mq}, any periodic orbit around the Schwarzschild and Kerr spacetimes can be indexed by the integers $(z,w,v)$. These three quantities have a geometric interpretation in terms of the structure of the trajectory, where $z$ is the `zoom' number, $w$ is the number of `whirls', and $v$ is the number of vertices formed by joining the successive apastra of the orbits \cite{Levin:2008mq}.

\begin{figure} 
 \begin{center}
 \begin{subfigure}[b]{0.45\textwidth}
   \centering
   \includegraphics[scale=0.455]{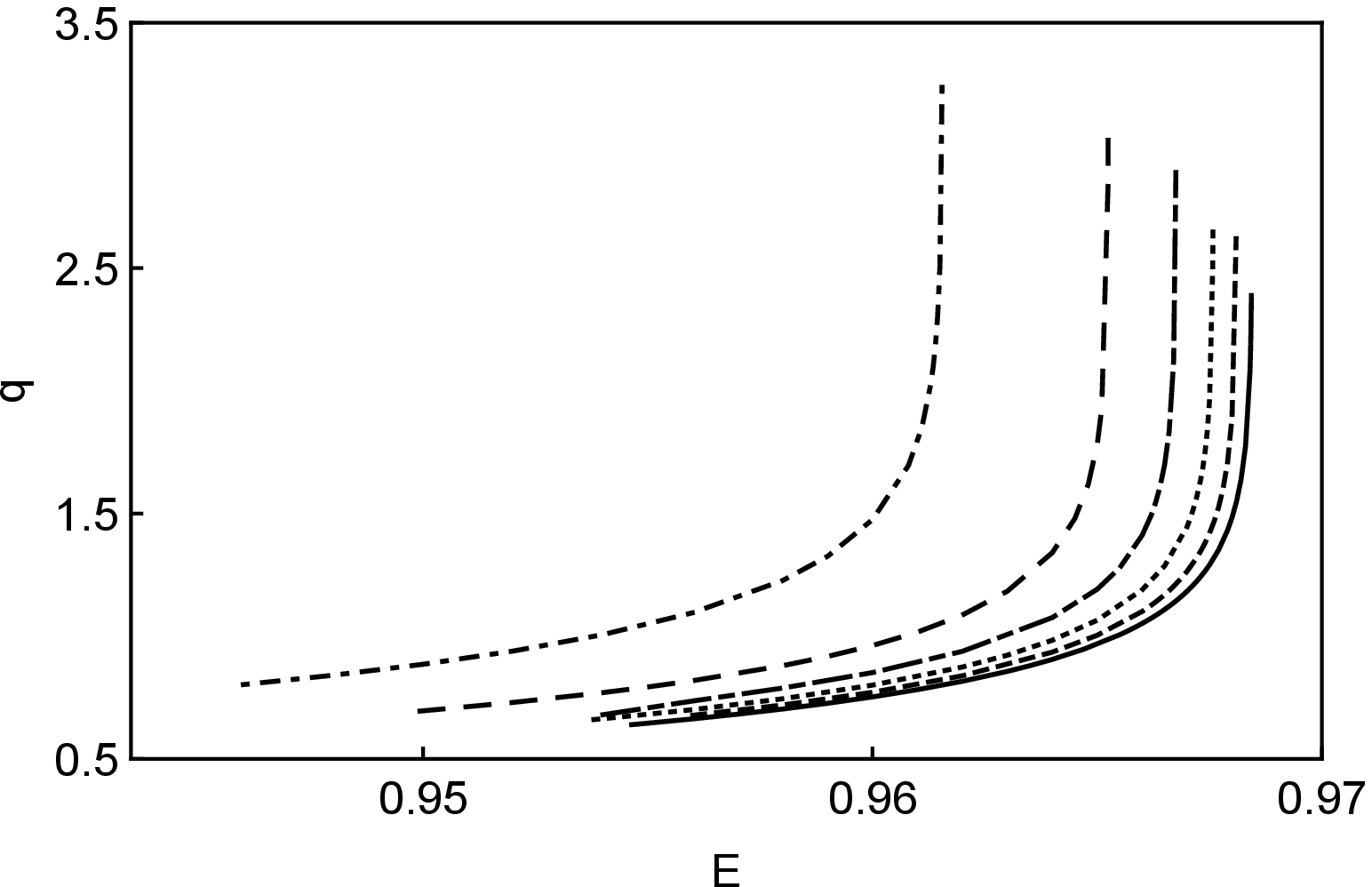}
  \end{subfigure}
  \begin{subfigure}[b]{0.45\textwidth}
   \centering
   \includegraphics[scale=0.45]{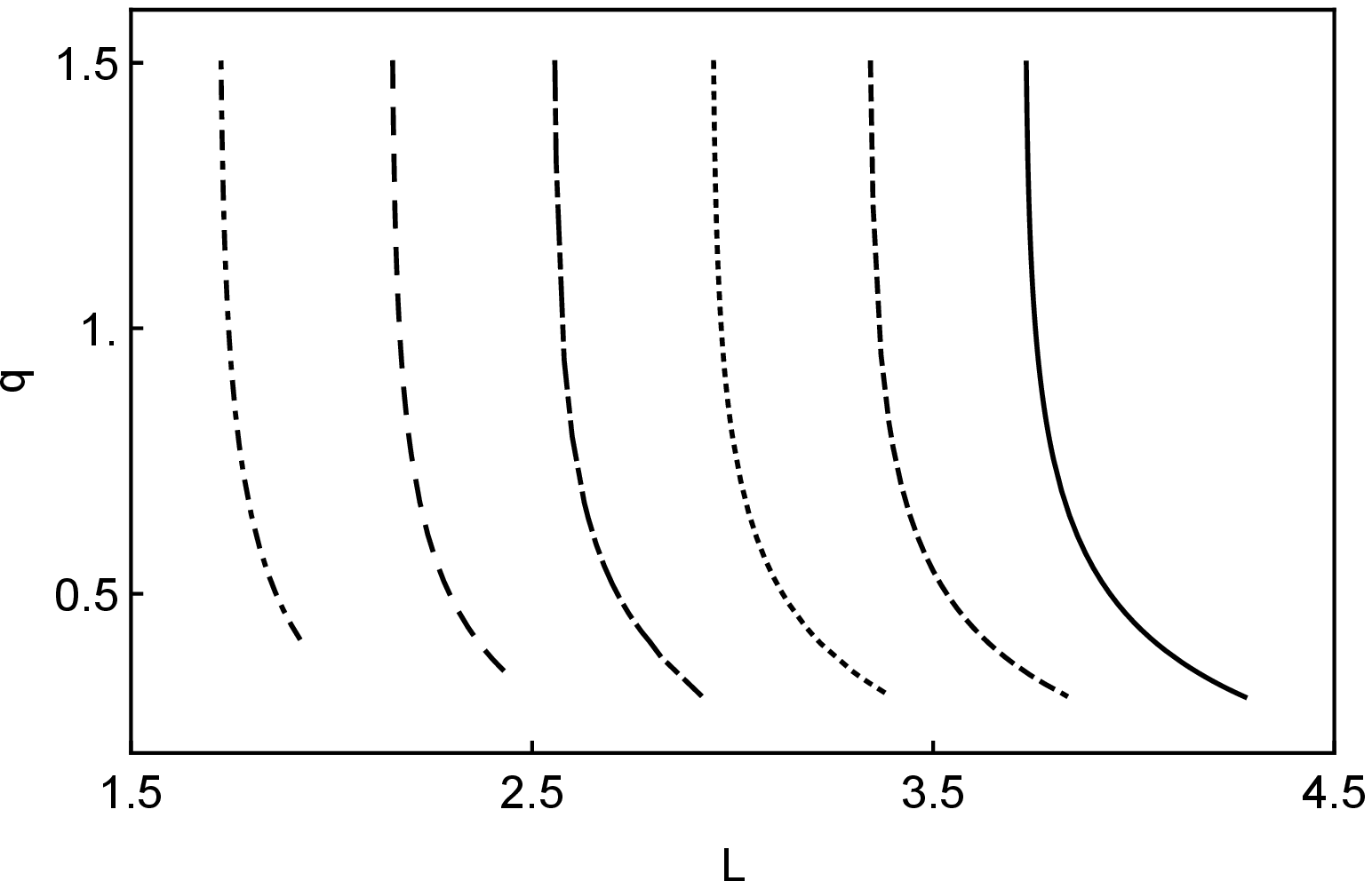}
  \end{subfigure}
  \end{center}
\caption{The variation of $q$ as a function of energy $E$ and angular momentum $L$ for different values of $\nu$: increasing from left to right the values are 0.5, 0.6, 0.7, 0.8, 0.9, and 1. In the left
plot, for each $\nu$ depicted above, the corresponding angular momentum is $L_\mathrm{av}$ as calculated from Eq.~\Eqref{L_av}. For the right plot energy is kept fixed for a (2,1,1) orbit. }\label{fig_qvsE}
\end{figure}

By seeking orbits with rational $q$ in the F/JNW spacetime for various $\nu$, we see the same periodic orbits that appear in the Schwarzschild cases, and thus the same three parameters $(z,w,v)$ to characterise the orbits. By comparing orbits of the same $(z,w,v)$ values for the Schwarzschild ($\nu=1$) and the F/JNW cases ($\nu<1$), we see that the energy of the F/JNW orbits are generally lower. Figure \ref{fig_qvsE} shows the variation of $E$ and $L$ with $q$. From the $q$ vs $E$ plot (the left-hand plot of Fig.~\ref{fig_qvsE}), we see that by adjusting the value of the energy for a given angular momentum, we may obtain a series of periodic orbits with rational $q$, ranging from stable-circular to homoclinic orbits. We observe that within each case, as $E$ is increased up to a certain value, $\Delta\phi$ (hence $q$) goes to infinity. The reason for this is similar to that observed in Ref.~\cite{Levin:2008mq}. When the energy increases up to that of a homoclinic orbit, the number of whirls approaches infinity, thus taking $\Delta\phi$ and $q$ to infinity as well. On the other hand, the stable circular orbit obviously corresponds to the case of minimum energy. It is also observed that as $\nu$ decreases from $1$ ($\nu=1$ being the Schwarzschild case), the same periodic orbit of a given $(z,w,v)$ requires lower energy. If we instead fix $E$ and vary $L$, as we see in the right-hand plot of Fig.~\ref{fig_qvsE}, the opposite behaviour occurs.

\begin{figure} 
 \begin{center}
 \begin{subfigure}[b]{0.25\textwidth}
   \centering
   \includegraphics[scale=0.25]{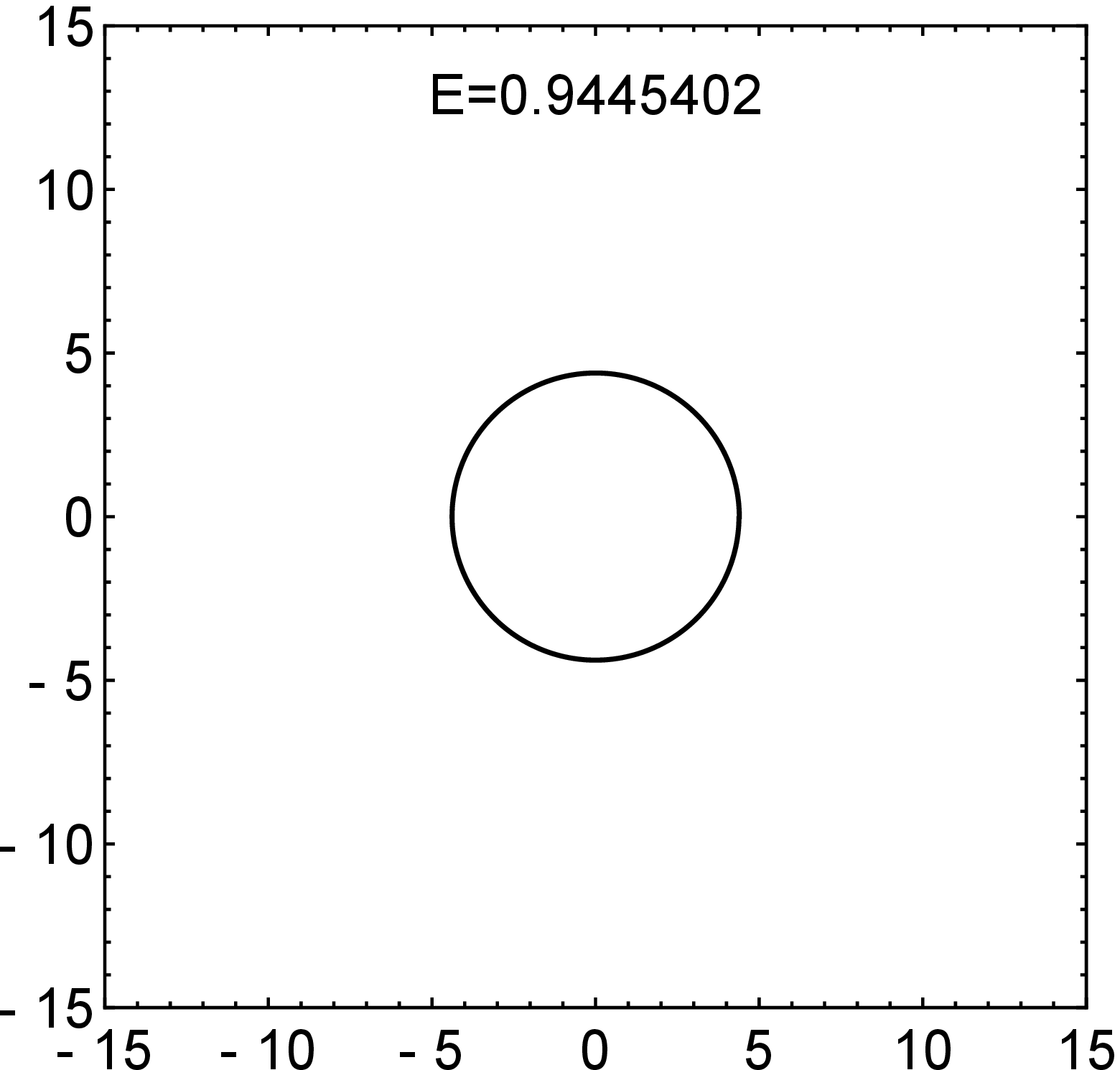} 
  \end{subfigure}
\begin{subfigure}[b]{0.25\textwidth}
   \centering
   \includegraphics[scale=0.25]{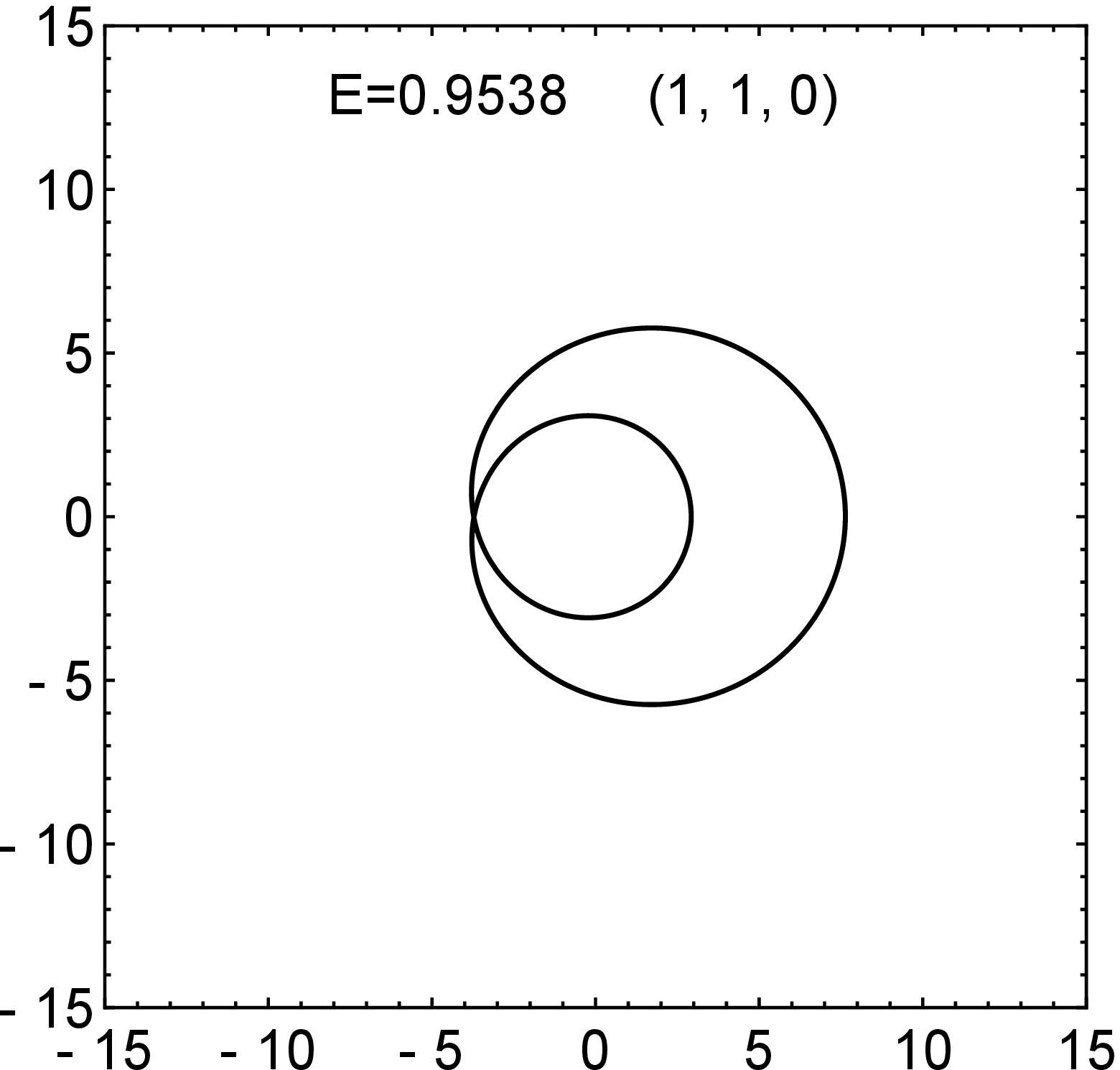} 
  \end{subfigure}
  \begin{subfigure}[b]{0.25\textwidth}
   \centering
   \includegraphics[scale=0.25]{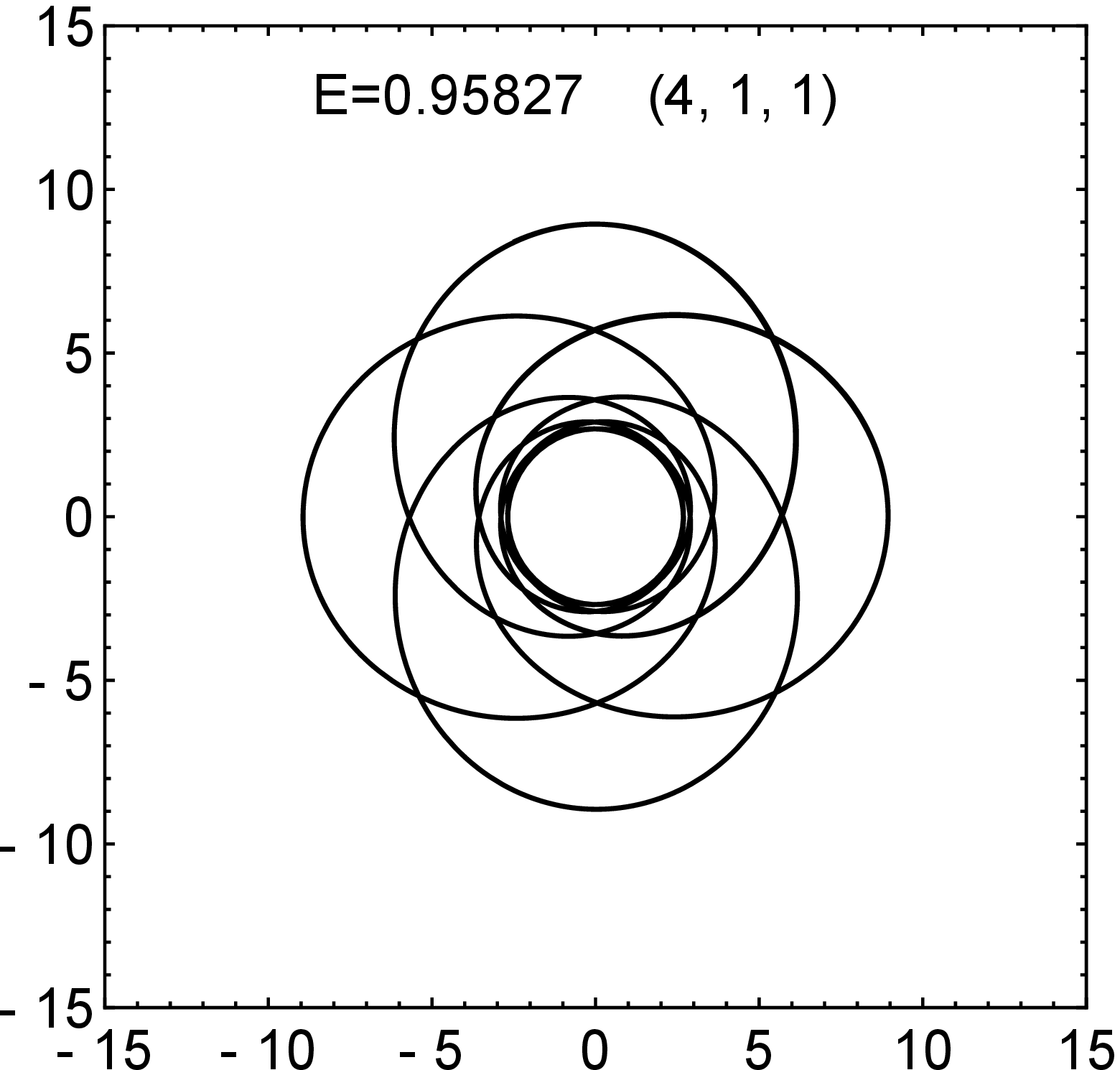} 
  \end{subfigure}
    \begin{subfigure}[b]{0.25\textwidth}
   \centering
   \includegraphics[scale=0.25]{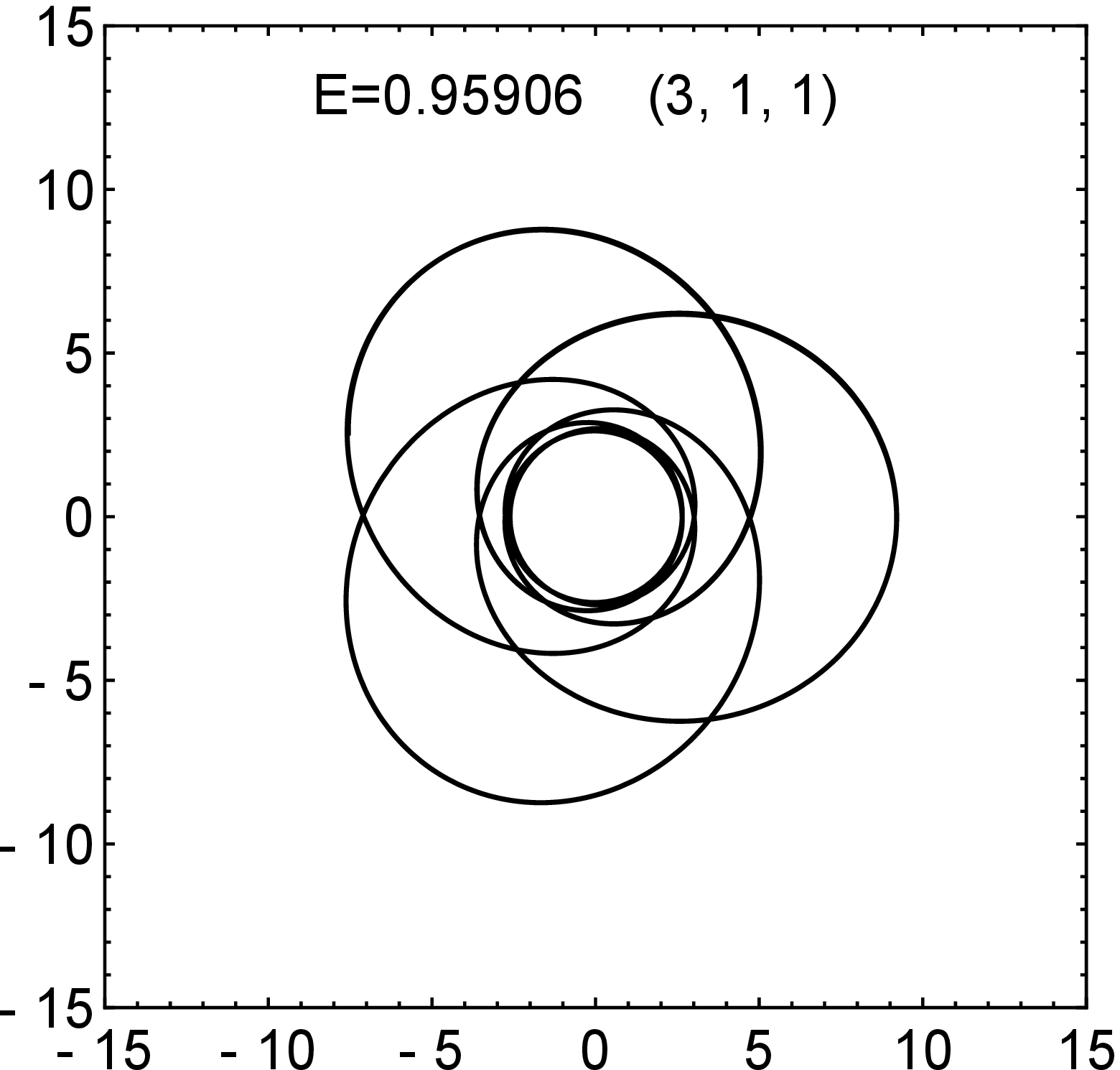} 
  \end{subfigure}
\begin{subfigure}[b]{0.25\textwidth}
   \centering
   \includegraphics[scale=0.25]{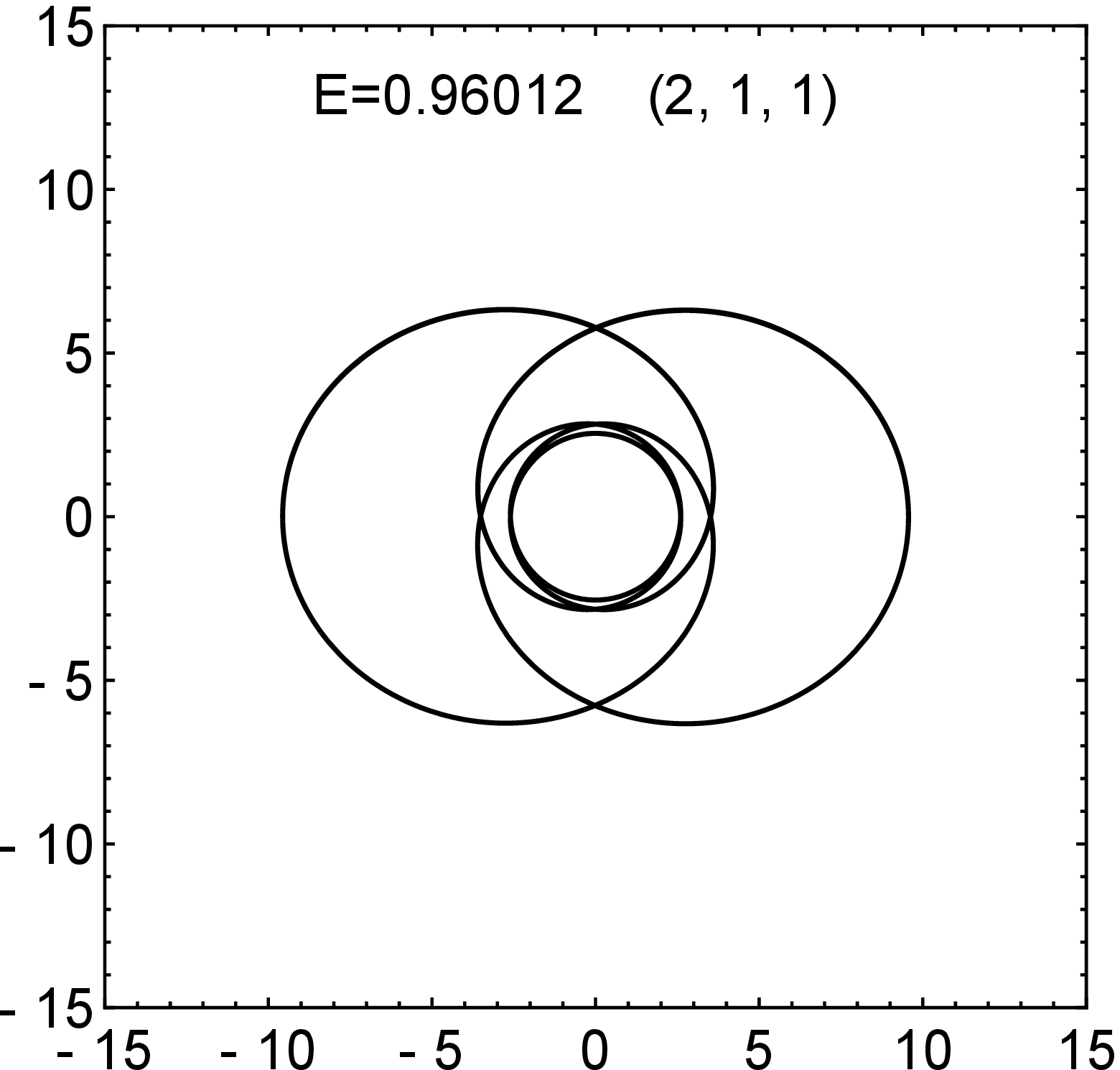} 
  \end{subfigure}
  \begin{subfigure}[b]{0.25\textwidth}
   \centering
   \includegraphics[scale=0.25]{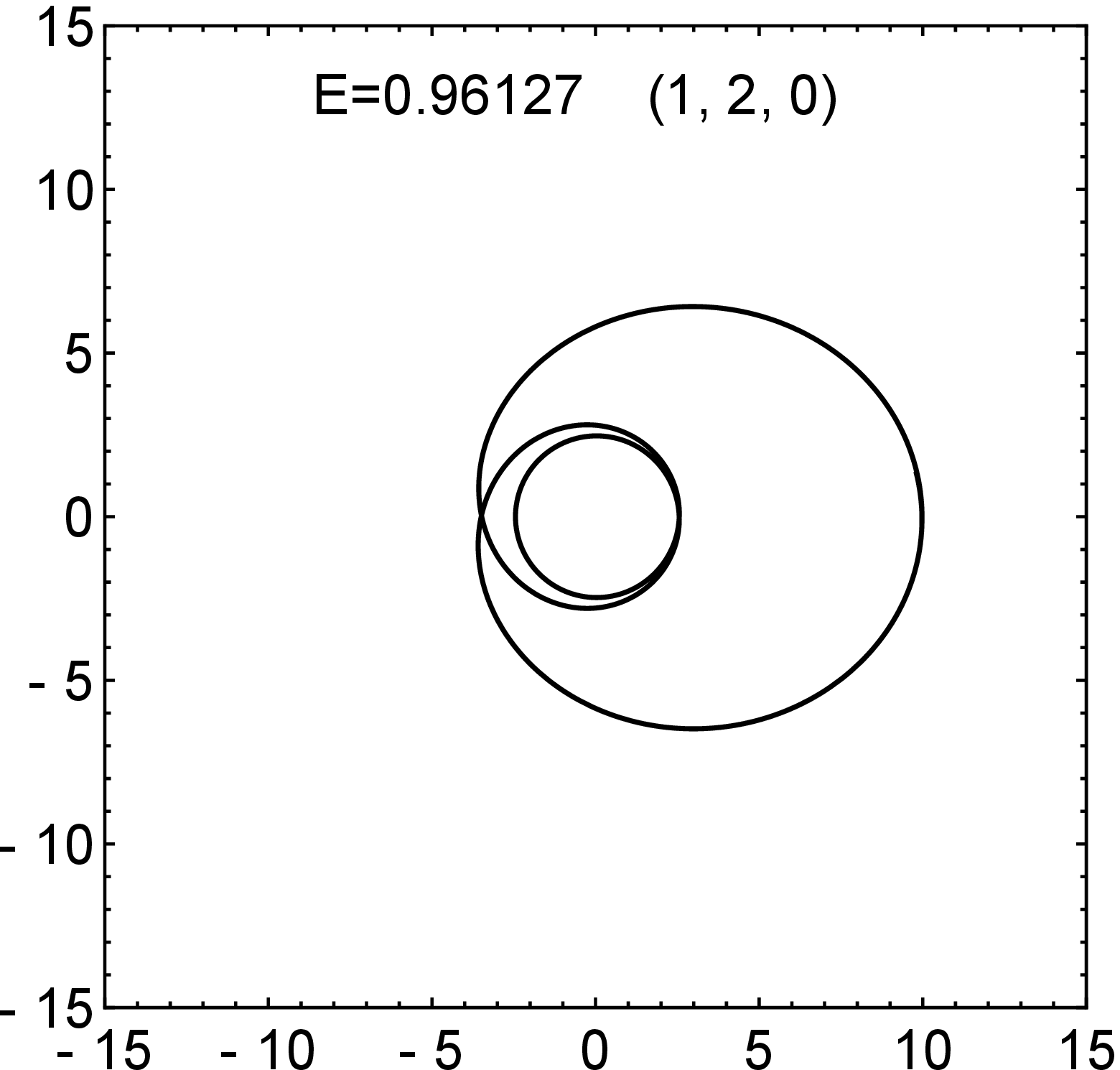} 
  \end{subfigure}
  \begin{subfigure}[b]{0.25\textwidth}
   \centering
   \includegraphics[scale=0.25]{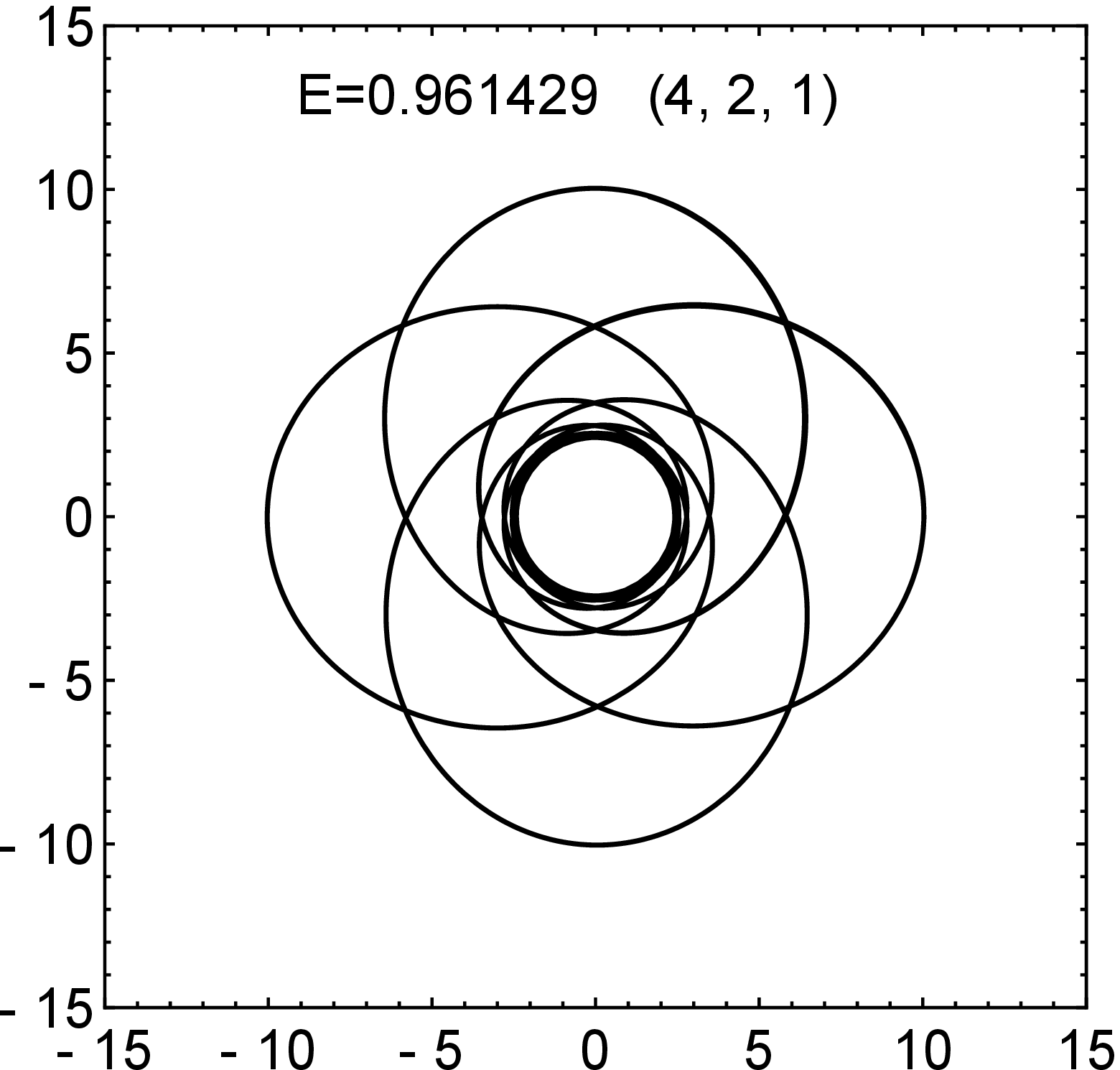} 
  \end{subfigure}
   \begin{subfigure}[b]{0.25\textwidth}
   \centering
   \includegraphics[scale=0.25]{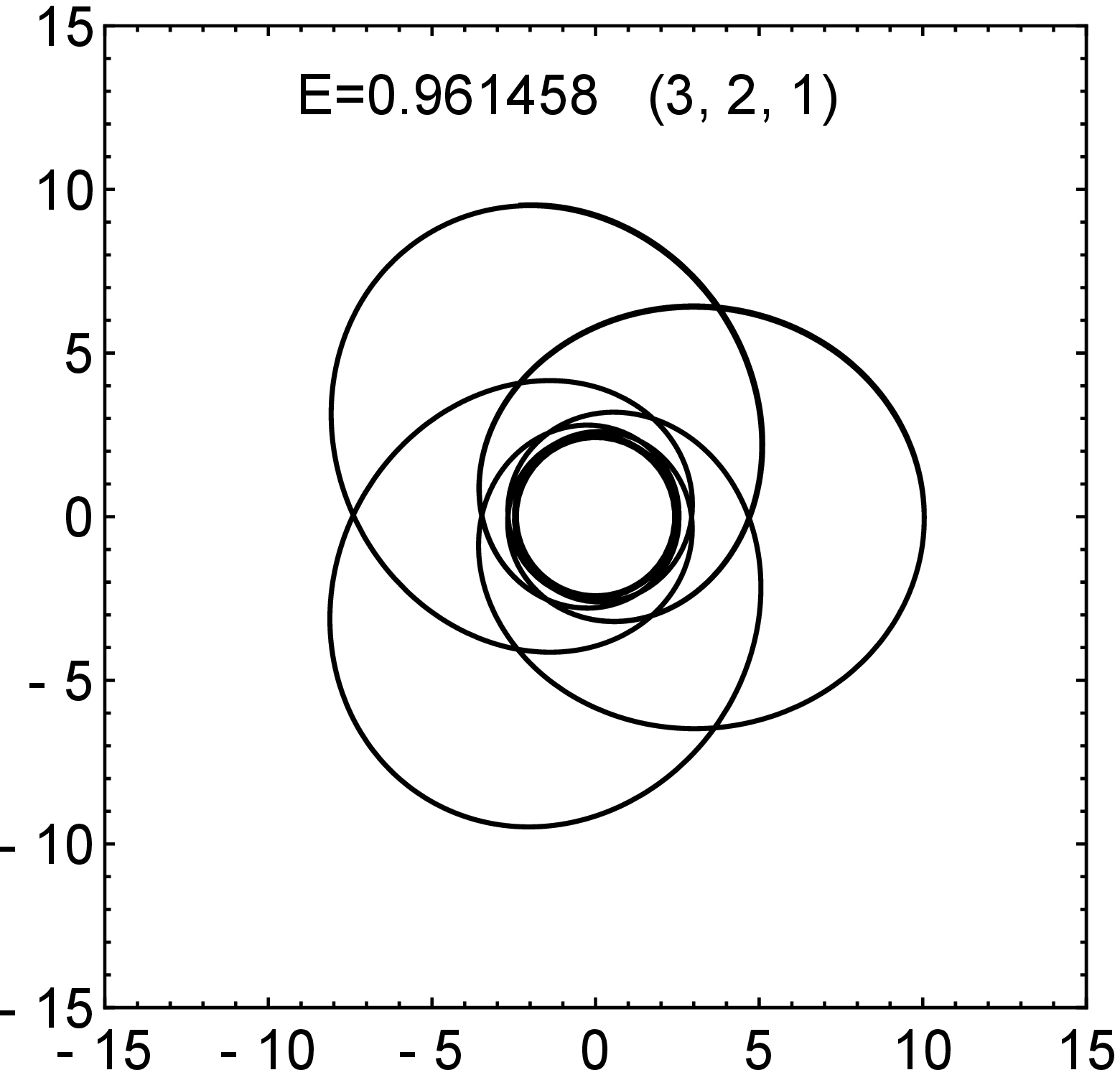} 
  \end{subfigure}
   \begin{subfigure}[b]{0.25\textwidth}
   \centering
   \includegraphics[scale=0.25]{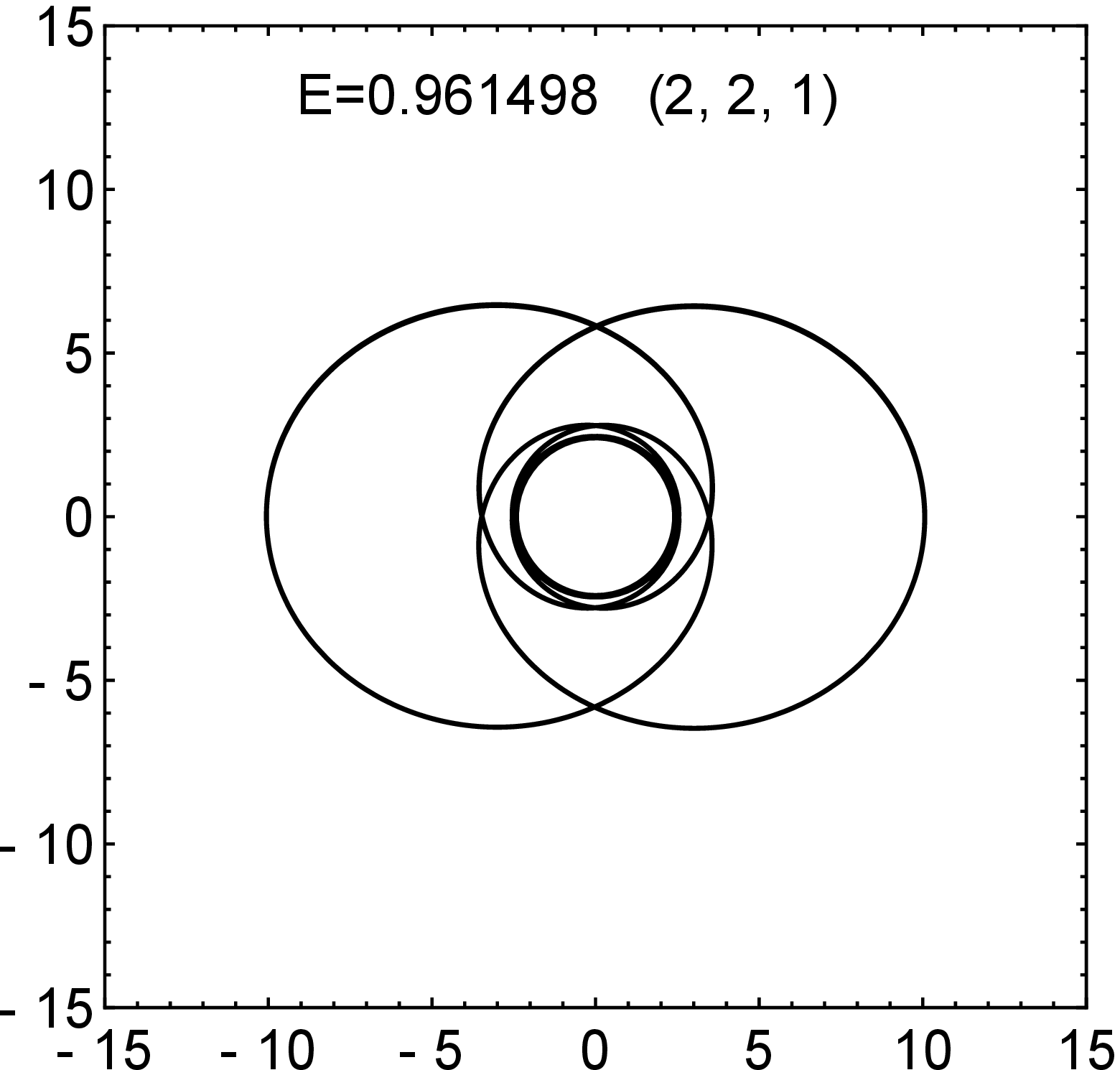} 
  \end{subfigure}
\end{center}
\caption{A series of $(z=1,2,3,4, w=1,2, v=1)$ orbits for $\nu=0.5$ with the angular momentum ascribed as $L_\textbf{av}$.}\label{Fige}
\end{figure}

Figure \ref{Fige} provides a visualisation of periodic orbits for $\nu=0.5$ with their corresponding energies. For a more systematic exploration, we confine our analysis to (1,1,0), (2,1,1), (3,1,1) and (4,1,1) orbits. From the data given in  Table \ref{tab1}, it is concluded that each specific orbit has a unique energy value which increases with the increasing $\nu$.


\begin{table}
 \begin{tabular}{ p{1cm} p{2.3cm} p{2.3cm} p{2.3cm} p{2.3cm} p{2.3cm}}
\hline
\textbf{$\nu$} & $L_{\mathrm{av}}$ &$E_{(1,1,0)}$  &$E_{(2,1,1)}$ &$E_{(3,1,1)}$ &$E_{(4,1,1)}$ \\
 \hline
0.5    &1.724485   &0.9538          &0.96012              &0.95906          &0.95827\\
0.6    &2.15245   &0.96074          &0.964555             &0.96397          &0.9635\\
0.7    &2.5568    &0.96305          &0.966215             &0.965734          &0.965356 \\
0.8    &2.95248   &0.96424          &0.967108             &0.96668          &0.966337\\
0.9    &3.34364   &0.96494          &0.967658             &0.967259          &0.966936\\
1      &3.732055    &0.96541        &0.968026             &0.967644          &0.967335\\
\hline
\end{tabular}
\caption{The energy values of $(z=1,2,3,4, w=1, v=1)$ orbits for various $\nu$ are presented with their corresponding angular momentum, $L_\textbf{av}$.}\label{tab1}
\end{table}

\section{Energy of generic orbits}\label{energy}

Transitions in the periodic orbits can be observed when the energy and angular momentum changes, which emanate in the form of gravitational waves. Hence, the rate of change of $q$ can be written in terms of the change of energy and angular momentum  \cite{Levin:2009sk}, i.e., $\frac{\dif q}{\dif t}=\frac{\partial q}{\partial E}\frac{\dif E}{\dif t}+\frac{\partial q}{\partial L}\frac{\dif L}{\dif t}$.

Figure \ref{fig_qvsE} shows $\frac{\dif q}{\dif E}>0$ and $\frac{\dif q}{\dif L}<0$, indicating that it is possible to satisfy the condition $\frac{\dif q}{\dif t}\approx0$, similar to the Schwarzschild and Kerr cases \cite{Levin:2009sk}. Hence, a test particle (perhaps a black hole) could merge into the naked singularity during an inspiral to produce gravitational waves in a process similar to the Schwarzschild and Kerr cases, albeit with different energy output. It is also observed that the increasing strength of the massless scalar field decreases the energy required for the same $q$.

\begin{figure} 
 \begin{center}
 \begin{subfigure}[b]{0.45\textwidth}
   \centering
   \includegraphics[scale=0.45]{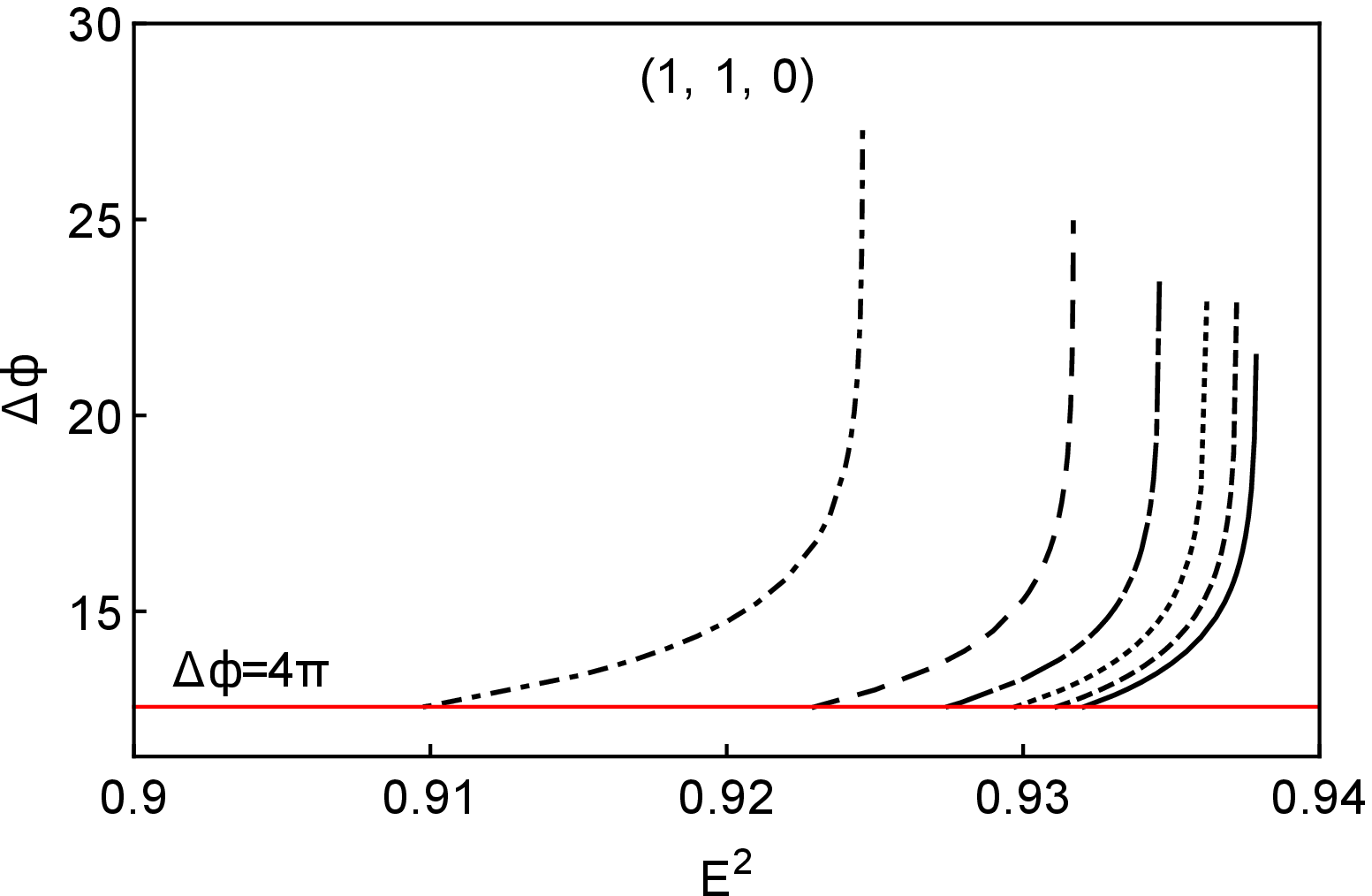}
  \end{subfigure}
\begin{subfigure}[b]{0.45\textwidth}
   \centering
   \includegraphics[scale=0.45]{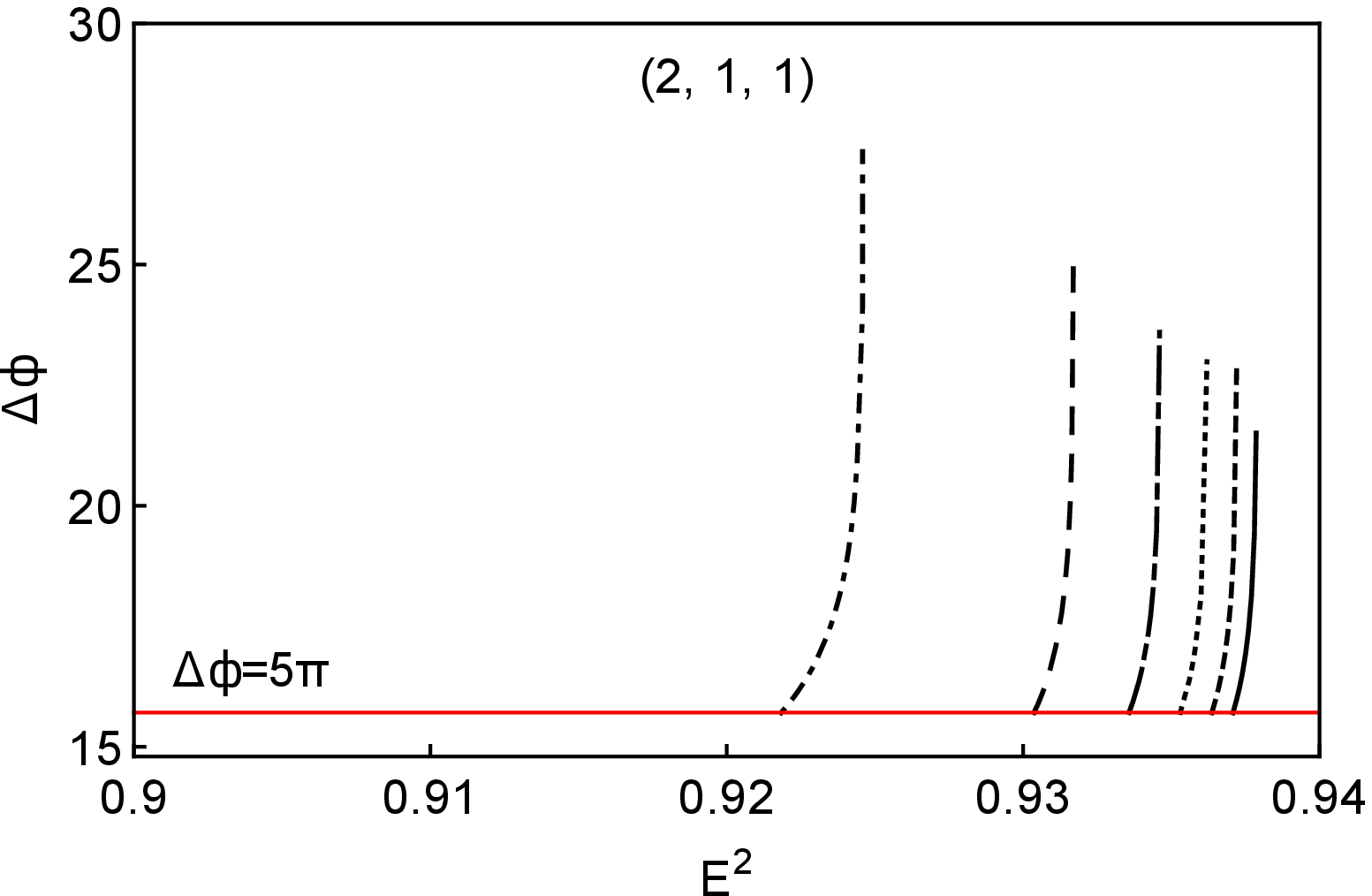}
  \end{subfigure}
  \begin{subfigure}[b]{0.45\textwidth}
   \centering
   \includegraphics[scale=0.45]{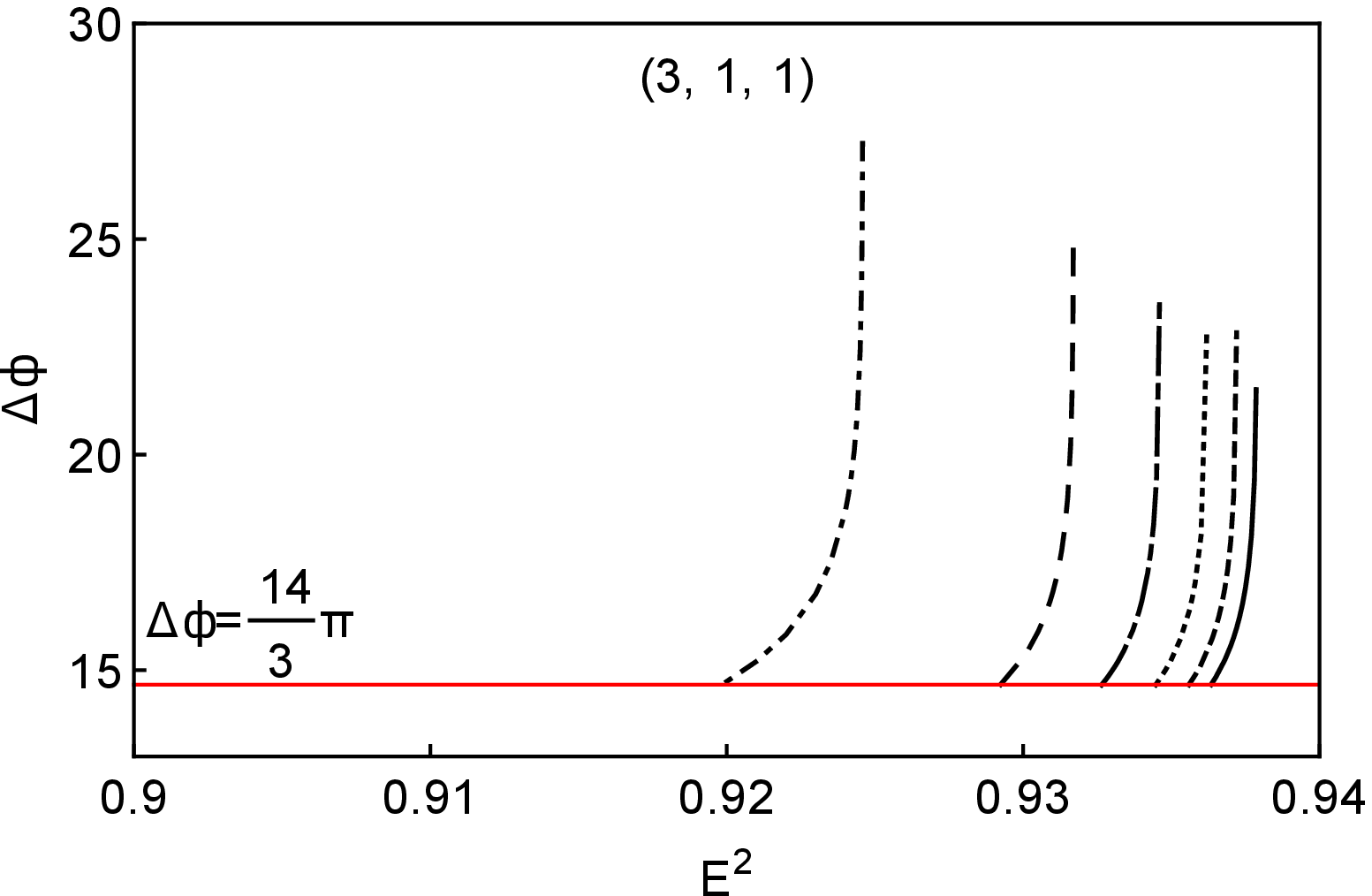}
  \end{subfigure}
   \begin{subfigure}[b]{0.45\textwidth}
   \centering
   \includegraphics[scale=0.45]{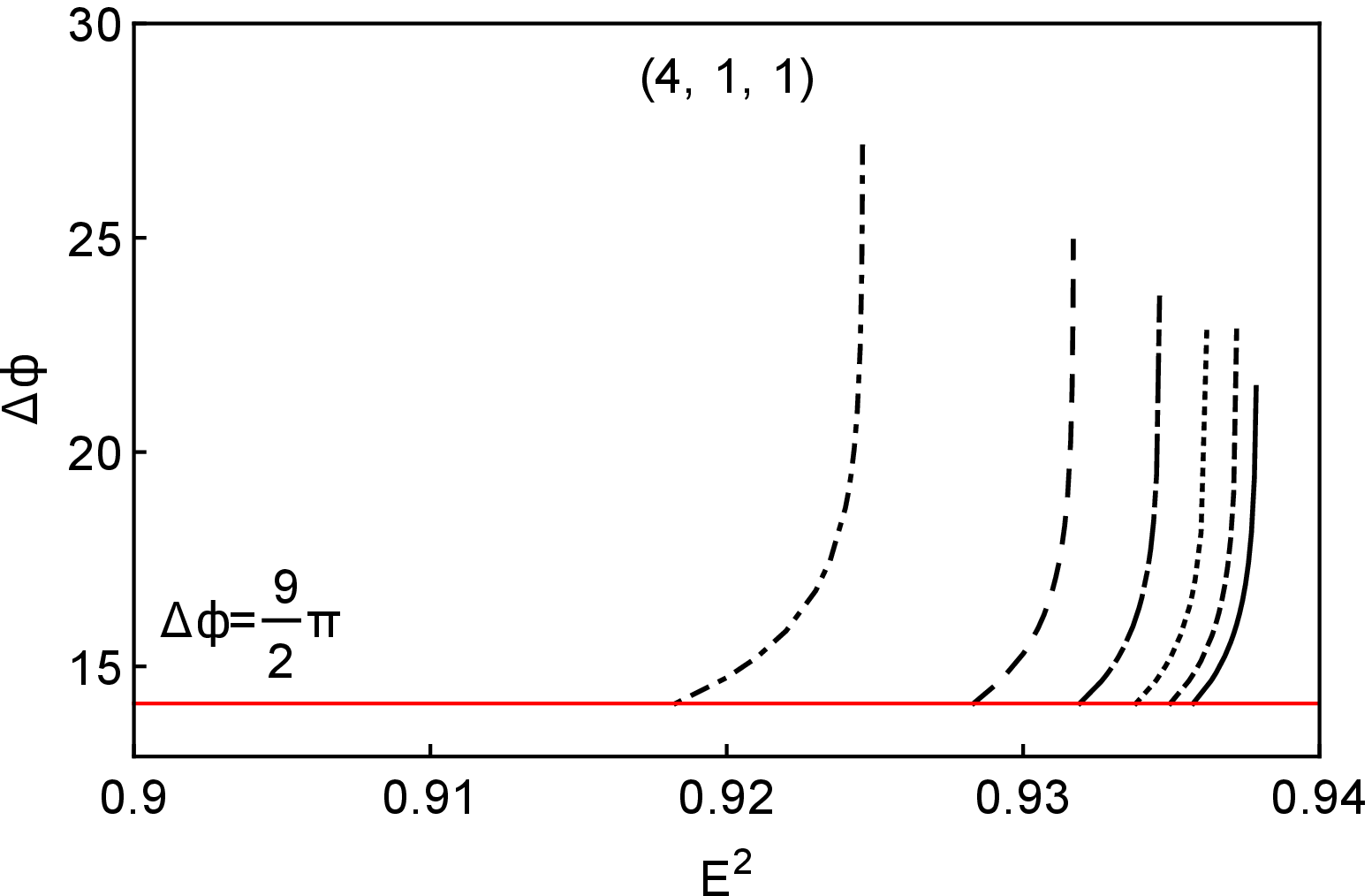}
  \end{subfigure}
  \end{center}
\caption{Plots of $\Delta\phi$ vs energy squared for ($z=1,2,3,4, w=1, v=1$) orbits with angular momenta $L = L_\mathrm{av}$. Increasing from left to right the values of $\nu$ are 0.5, 0.6, 0.7, 0.8, 0.9 and 1. The horizontal line corresponds to the exact periodic orbits where $\Delta\phi=4\pi$, $5\pi$, $\frac{14\pi}{3}$, and $\frac{9\pi}{2}$, respectively.}\label{Figf}
\end{figure}

Having established the qualitative features of orbits in the F/JNW spacetime, we now turn to consider the observational features that might distinguish it from the trajectory around a Schwarzschild black hole. To do this we consider the apastron shift of quasi-periodic orbits. Particularly, since every quasi-periodic orbit can be considered a perturbation of a periodic orbit, we can measure how `far' a given quasi-periodic orbit is from its periodic counterpart by considering the accumulated angle $\Delta\phi$ as defined in Eq.~\Eqref{DeltaPhiIntegral}.

The behaviour of  $\Delta\phi$ when the energy of each orbit increases away from the periodic orbit values is illustrated in Fig.~\ref{Figf}. From the figure, we may deduce that, regardless of any orbit considered, the spacetime with decreasing $\nu$ (i.e., stronger scalar field) shows a higher precession as we move away from the periodic orbits.

\section{Conclusion} \label{conclusion}

By using the Hamiltonian formulation, we have analysed the geodesic motion of a time-like particle in the F/JNW spacetime. This spacetime corresponds to a spherically symmetric solution sourced by a massless scalar field. By analysing the effective potential of the particle, we have recovered the earlier results of \cite{Chowdhury:2011aa,Babar:2015kaa} where for $\nu<\half$, any particle with non-zero angular momentum encounters a potential barrier which prevents it from reaching the naked singularity $r=r_g$.

To reduce the number of free parameters, we have defined a specific angular momentum $L_{\mathrm{av}}$ for any given $\nu$ and $r_g$ as the average of the angular momenta of the marginally stable and marginally bound circular orbits. This definition is somewhat arbitrary, though we have orbits with angular momentum $L_{\mathrm{av}}$ which gives a sufficiently deep potential well that allows a rich variety of bound orbits. The periodic orbits in the F/JNW spacetime are qualitatively similar to those in the Schwarzschild and Kerr spacetimes, though for the same orbits of a given $(z,w,v)$ the required energy is lower for a spacetime with a stronger scalar field strength. Orbits with generic, non-integer $q$ have also been considered. In particular, we found that the precession $\Delta\phi$ increases in the presence of the massless field compared to the Schwarzschild case. These may provide a possible observational signature by observing trajectories around a central source to distinguish a naked singularity from a black hole. Decaying orbits with non-trivial $\frac{\dif q}{\dif t}$ may also be observationally relevant, in light of the recent detection of gravitational waves by LIGO. This provides a complementary observational avenue to other candidates, for instance, detection via gravitational lensing \cite{DeAndrea:2014ova}.

\bibliographystyle{jnwperiod}

\bibliography{jnwperiod}

\end{document}